\documentclass[aps,floatfix,prb,twoside,twocolumn,10pt,showpacs,citeautoscript,superscriptaddress]{revtex4-2}

\usepackage{amsmath}
\usepackage{amssymb}
\usepackage{float}
\usepackage{glossaries}
\usepackage{graphicx}
\usepackage[version=4]{mhchem}
\usepackage{siunitx}
\usepackage{tabularx}
\usepackage[dvipsnames]{xcolor}
\usepackage[unicode, colorlinks=true, urlcolor=blue, linkcolor=blue,    citecolor=blue]{hyperref}

\hypersetup{pdfauthor={C. Linderälv, J. M. Rahm, and P. Erhart}}
\hypersetup{pdftitle={High-throughput characterization of transition metal dichalcogenide alloys: Thermodynamic stability and electronic band alignment}}

\DeclareSIUnit{\formulaunit}{f.u.}

\setacronymstyle{long-short}
\newacronym{2d}{2D}{two-dimensional}
\newacronym{bec}{BEC}{Born effective charge}
\newacronym{cbm}{CBM}{conduction band minimum}
\newacronym{ce}{CE}{cluster expansion}
\newacronym{dft}{DFT}{density functional theory}
\newacronym{mc}{MC}{Monte Carlo}
\newacronym{rmse}{RMSE}{root mean square error}
\newacronym{si}{SI}{Supplementary Information}
\newacronym{sisso}{SISSO}{sure independence screening and sparsifying operator}
\newacronym{sqs}{SQS}{special quasi-random structure}
\newacronym{tmd}{TMD}{transition metal dichalcogenide}
\newacronym{vbm}{VBM}{valence band maximum}
\newacronym{xc}{XC}{exchange-correlation}

\global\let\oldnewlabel\newlabel
\gdef\newlabel#1#2{\newlabelxx{#1}#2}
\gdef\newlabelxx#1#2#3#4#5#6{\oldnewlabel{#1}{{#2}{#3}}}
\let\newlabel\oldnewlabel
\begin{document}

\title{
    High-throughput characterization of transition metal dichalcogenide alloys:
    Thermodynamic stability and electronic band alignment
}
\author{Christopher Linder\"{a}lv}
\author{J. Magnus Rahm}
\author{Paul Erhart}
\affiliation{Department of Physics, Chalmers University of Technology, SE-41296, Gothenburg, Sweden}
\email{erhart@chalmers.se}

\begin{abstract}
Alloying offers a way to tune many of the properties of the \gls{tmd} monolayers. 
While these systems in many cases have been thoroughly investigated previously, the fundamental understanding of critical temperatures, phase diagrams and band edge alignment is still incomplete. 
Based on first principles calculations and alloy cluster expansions we compute the phase diagrams 72 \gls{tmd} monolayer alloys and classify the mixing behaviour. 
We show that ordered phases in general are absent at room temperature but that there exists some alloys, which have a stable Janus phase at room temperature. Furthermore, for a subset of these alloys, we quantify the band edge bowing and show that the band edge positions for the mixing alloys can be continuously tuned in the range set by the boundary phases.
\end{abstract}
\maketitle
\section{Introduction}

\begin{figure*}[bth]
    \centering
    \includegraphics{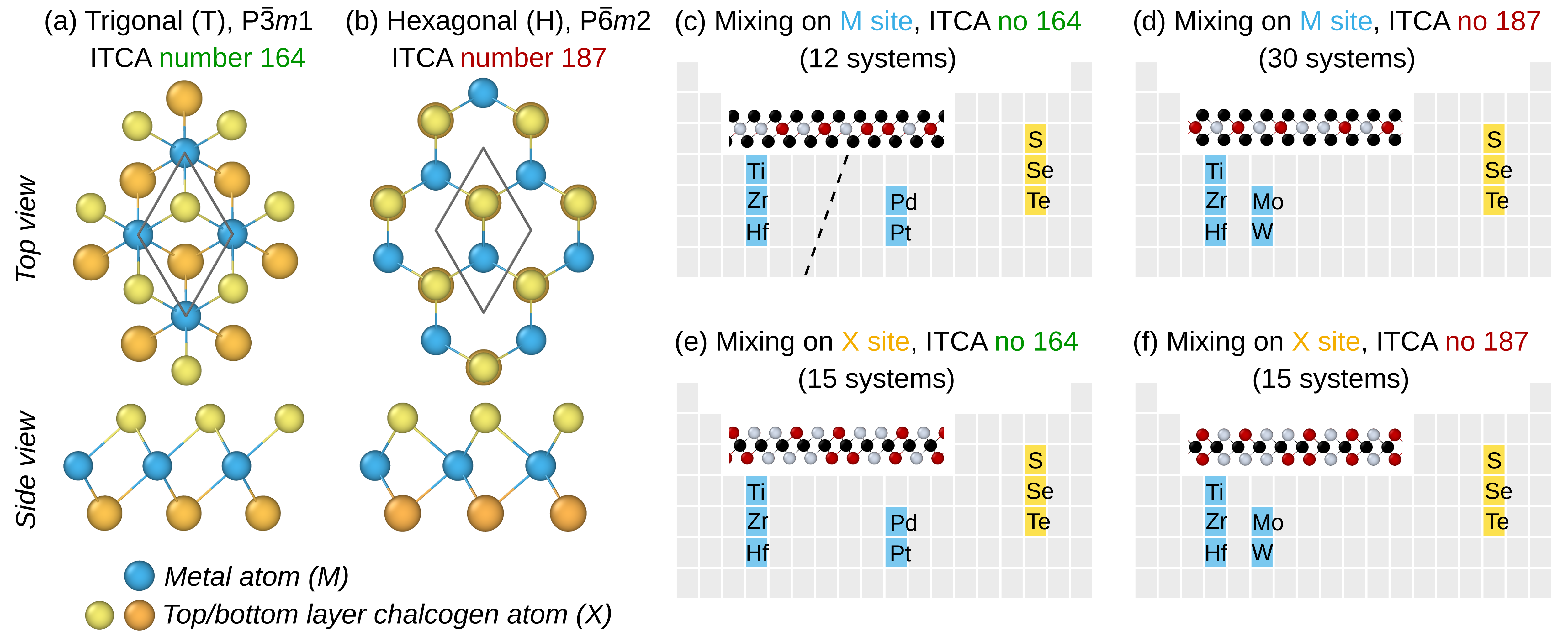}
    \caption{
    Overview of the 72 \gls{tmd} monolayer alloys considered in this work in (a)~space group P$\bar{3}m1$ (ITCA number 164) and (b)~space group P$\bar{6}m2$ (ITCA number 187).
    The alloys consisted of either (c--d) two species on the metal (M) site and one (no alloying) on the chalcogen (X) site, or (e--f) one species (no alloying) on the M site and two species on the X site.
    For space group 187, we considered Ti, Zr, Hf, Pd, and Pt on the M site, but we never mixed Ti/Zr/Hf with Pd/Pt.
    For space group 164, we considered Ti, Zr, Hf, Mo, and W.
    The X site was always occupied by S, Se, or Te, either just one of them (M-site mixing), or a mixture of two of them (X-site mixing).}
    \label{fig:overview}
\end{figure*}

Monolayer \glspl{tmd} constitute a class of \gls{2d} materials, commonly of \ce{MX2} stoichiometry, where M is a transition metal (e.g., Mo, W, Zr, Hf, Ti, Pd, Pt) and X is a chalcogen (S, Se, Te). 
These monolayer compounds exhibit a structure where the transition metal is sandwiched between chalcogen atoms and exhibit hexagonal or trigonal symmetry (\autoref{fig:overview}a).
\Glspl{tmd} have received a considerable amount of attention over the last decade, e.g., due to the excellent optical properties of the semiconducting group 6 \gls{tmd} \ce{MoS2} \cite{MakLeeHon10}, the prospects of vertical integration into heterostructures \cite{AndLatThy15}, the emergent properties of moiré structures \cite{NaiJai18, BreLinErh20}, and the outlook of using \glspl{tmd} as catalysts for the hydrogen evolution reaction \cite{HinMosBon05}.
Many \gls{tmd} monolayers exhibit moderate band gaps of around 1 to \SI{3}{\electronvolt}, which is suitable for photochemical applications or for use as channel materials in nanoelectronics \cite{YooGanSal11, RadRadBri11, YaoYan22}. 
Due to the emergence of high-$\kappa$ dielectrics as gate material in transistors such as \ce{ZrO2} and \ce{HfO2}, \gls{2d} \glspl{tmd} based on the same transition metals may also play a role in future high-$\kappa$ transistor designs \cite{MleZhaLee17}.

The electrical transport properties as well as the optical response is ultimately dependent on band edge positions. 
The position of the band edges are particularly important in devices based on heterojunctions.
Specifically, \gls{tmd} heterostructures with type-II alignment can be used for photoseparation of charge carriers \cite{WanLiChe19}.
There are various ways to tailor the electronic properties, including functionalization \cite{TanJia15}, Coulomb engineering \cite{RajChaYu17, PadPeeJan14, RiiManThy20}, strain engineering \cite{ZhaLiTer18, ConWanZie13}, and alloying \cite{RiiManThy20, CheXiDum13, KomKra12}.
While alloying may allow for a continuous tuning of the band edge position, not all alloys can be manufactured due to their inherent thermodynamical properties.
The main objectives of this work are therefore to map out the phase diagrams of a large number of binary \gls{tmd} alloys and the band edge variations of promising alloy combinations.

All of the binary alloys based on \ce{(Mo, W)(S,Se)2} have been synthesized \cite{Xie15}, usually in the hexagonal H structure (spacegroup P$\bar{6}m2$, ITCA number 187; \autoref{fig:overview}b), and have been thoroughly studied in the context of band gap engineering by composition control \cite{KomKra12, KanTonLi13, KutPenYak14, YanYak18, TanWeiLiu2017, DuaWanFan16, WanLiuBas20}.
It has been shown that these alloys are random at ambient conditions due to (very) small mixing energies \cite{KomKra12}.
Although ordered phases have been found in computational studies \cite{Xie15}, these are not observed in experiments \cite{DumKobLiu13, XiaLohVin21}, most likely due to the very small energy gains on the order of \SI{1}{\milli\electronvolt} compared to the respective disordered solution.
In addition, the \ce{MoS_{2x}Te_{2(1-x)}} alloy has been fabricated \cite{AptKriHac18}.
For the optical properties, in particular, it is relevant whether alloying preserves the valley degree of freedom found in the boundary phases, i.e., the non-alloyed systems.
In this regard, it is noteworthy that, e.g., \ce{Mo_{x}W_{1-x}S2} has been recently shown to be robust in this regard \cite{XiaLohVin21}. 

While the group 6 \gls{tmd} alloys have been extensively studied, the prospect of alloying these \glspl{tmd} with transition metals from other groups of the periodic table has not been systematically addressed, although doping of e.g., \ce{MoS2} with Ti has been reported \cite{TedLewObr16}.
In addition, there are various families of \glspl{tmd}, including different structural prototypes, for which the alloying behaviour is largely unexplored.
The \gls{tmd} monolayers based on Zr, Hf, Pd, and Pt preferentially crystallize in the trigonal T phase (space group P$\bar{3}m1$, ITCA number 164; \autoref{fig:overview}a) \cite{ZhaZhuWan15, ZhaQiaYu16, HaaMikMoh18, GjeTagRas21} and exhibit indirect band gaps.
The basic electronic structure and excitation spectra of Zr and Hf based boundary phase \glspl{tmd} have been studied previously \cite{LauCocDra19}.
Some theoretical \cite{OliFoxHas20} and experimental \cite{GaiZanKra04, MouZanJan09} studies on the electronic properties of the layered bulk alloys exist but to the best of our knowledge the thermodynamic and electronic properties of T-\glspl{tmd} monolayer alloys have not been systematically addressed.

A special subclass of monolayer alloys are Janus monolayers, i.e. ordered MXX' compounds where the chalcogens X and X' occupy the top and bottom layer sides of the \gls{tmd} sheet.
Janus monolayers have been suggested to be useful for band alignment engineering \cite{RiiManThy20} and as catalysts for the hydrogen evolution reaction \cite{ZhaJiaKho17}.
While some Janus structures have been manufactured \cite{LuZhuXia17}, little is known about the general thermodynamical stability of these phases.

In this work, we aim to provide a more comprehensive perspective of \gls{tmd} monolayer alloys.
We consider 72 binary alloys of the T and H structure types (\autoref{fig:overview}) and focus on their thermodynamic properties, specifically the underlying phase diagrams, and electronic properties, specifically band edge positions (for the semiconducting systems) and work functions (for the metallic systems).
In the next section we outline the methodology used in this work with details provided in the \gls{si}.
This is followed by a description and discussion of the phase diagrams and mixing characteristics in \autoref{sect:phase-diagrams}.
Models for the critical temperature and the structural categorization are presented in \autoref{sect:models} while the computed band edge positions and bowing parameters are summarized in \autoref{sect:electronic-structure}.

\section{Methodology}

Here, we study the thermodynamic and electronic properties of \gls{tmd} alloys with chemical formula \ce{MX2} (\autoref{fig:overview}) in the T structure (space group P$\bar{3}m1$, ITCA number 164) and the H structure (space group P$\bar{6}m2$; ITCA number 187).
For the T structure, we consider mixing on the M-site involving Ti, Zr, and Hf (9 systems) and Pd and Pt (3 systems) as well as X-site mixing (15 systems).
For the H structure, we consider mixing on the M-site involving, Ti, Zr, Hf, Mo, and W (30 systems) as well as X-site mixing (15 systems).
These alloys were selected based on the known stability or metastability of the boundary phases as documented in the Computational 2D Materials Database \cite{HaaMikMoh18}.

To assess the thermodynamic properties of these \gls{tmd} alloys, we constructed alloy \glspl{ce} \cite{AngMunRah19, scikit-learn} (\autoref{snote:cluster-expansion-construction}) using reference data from \gls{dft} calculations \cite{Blo94, KreHaf93, KreFur96, DioRydSch04, KliBowMic11, LinErh16, HarFor08, AngMunRah19} (\autoref{snote:dft-calculations-for-ce-construction}).
These models were subsequently sampled using \gls{mc} simulations \cite{AngMunRah19, Cow50} to obtain phase diagrams and critical temperatures (\autoref{snote:calculation-of-critical-temperatures}).

We then carried out a further characterization of the electronic properties of these systems, specifically focusing on the band gap and the position of the conduction and valence band edges using \gls{dft} calculations and \glspl{sqs} to mimic complete (fully random) mixing \cite{ZunWeiFer90, WalTiwJon13, AngMunRah19, Blo94, KreHaf93, KreFur96, DioRydSch04, KliBowMic11, BerHyl14, HeyScuErn03, HeyScuErn06} (\autoref{snote:dft-calculations-for-band-structure}).
From the latter analysis, we excluded alloy systems with very high critical temperatures, leaving us with 21 and 27 systems in structure types T and H, respectively.

\begin{figure*}[bht]
    \centering
    \includegraphics[width = 0.95\linewidth]{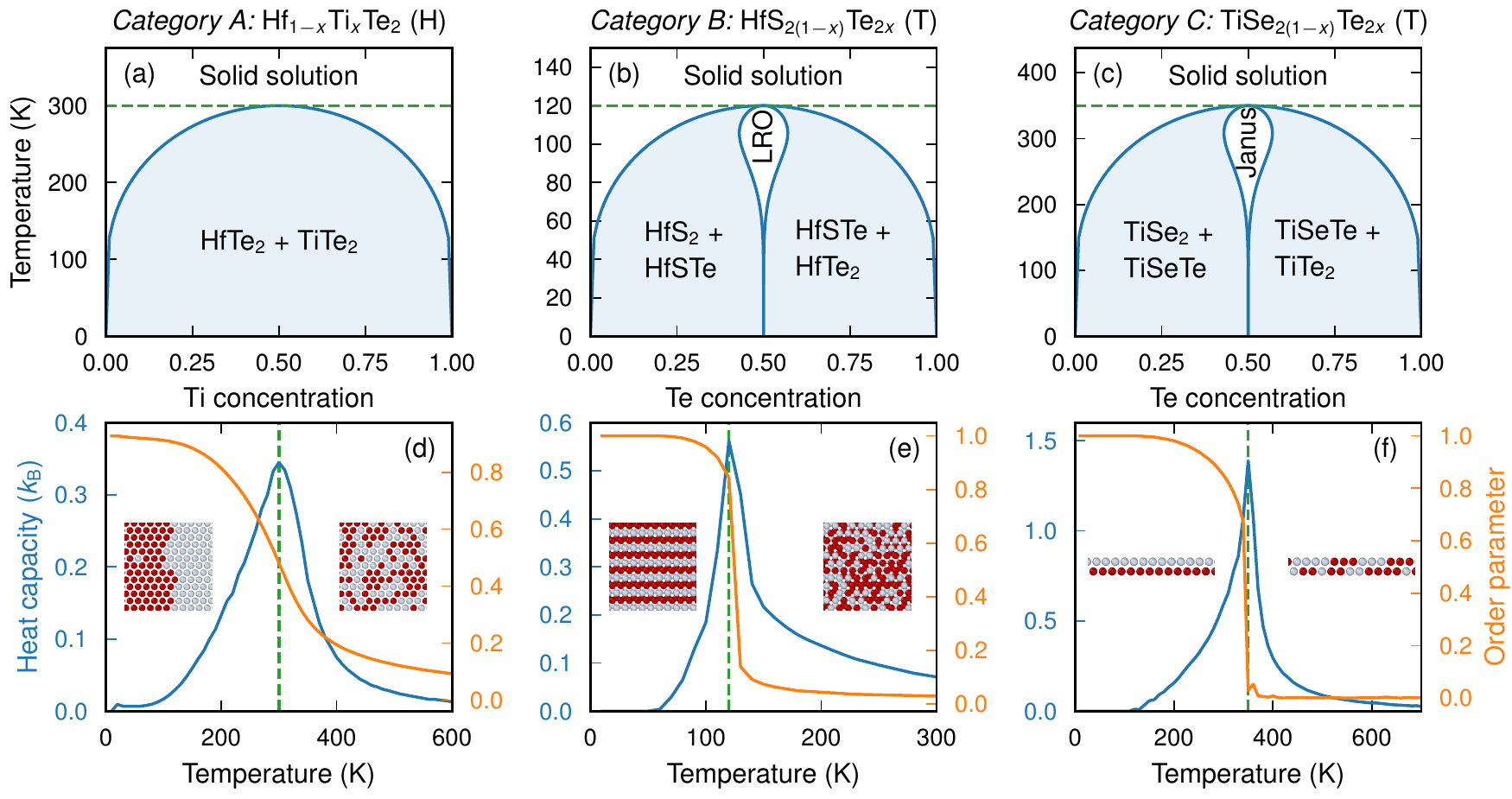}
    \caption{
    Schematic phase diagrams of systems that show (a) phase separation (category A), (b) an intermediate phase at 50\% with long-range in-plane order (category B), and (c) an intermediate phase at 50\% with long-range out-of-plane (or Janus-type) order (category C).
    (d--f)~The critical temperatures (dotted, green lines) are identified as the position of the peak in heat capacity (blue lines) as calculated with Monte Carlo simulations at 50\% concentration.
    The peaks correspond to a rapid decline in the relevant order parameter (orange lines), here chosen as (d) the first-nearest neighbor short-range order, (e) the in-plane structure factor, and (f) the difference in concentration between the two layers (for details, see \autoref{snote:calculation-of-critical-temperatures}).
    Insets show snapshots of small parts of the structures in the \gls{mc} simulations (top view in (d) and (e), side view in (f)).}
    \label{fig:order-parameters-examples}
\end{figure*}

\section{Results and discussion}

\subsection{Thermodynamic properties}
\label{sect:phase-diagrams}

The mixing behavior at zero temperature can be grouped into three different categories:
(A) systems that exhibit two boundary phases separated by a miscibility gap and no long-range ordered mixed phase ({\autoref{fig:order-parameters-examples}a}) as well as systems that exhibit a long-range ordered phase at 50\% composition with (B) in-plane order ({\autoref{fig:order-parameters-examples}b}) or (C) out-of-plane order ({\autoref{fig:order-parameters-examples}c}), a situation that is commonly referred as a Janus system \cite{LuZhuXia17}.

Out of the 72 alloys considered, we find that 49 alloys fall into category A (\autoref{sect:category-a}), 12 into category B (\autoref{sect:category-b}), and 11 into category C (\autoref{sect:category-c}).
These categories are discussed in more detail in the following sections.

\begin{figure*}[htb]
    \centering
    \includegraphics[width=\linewidth]{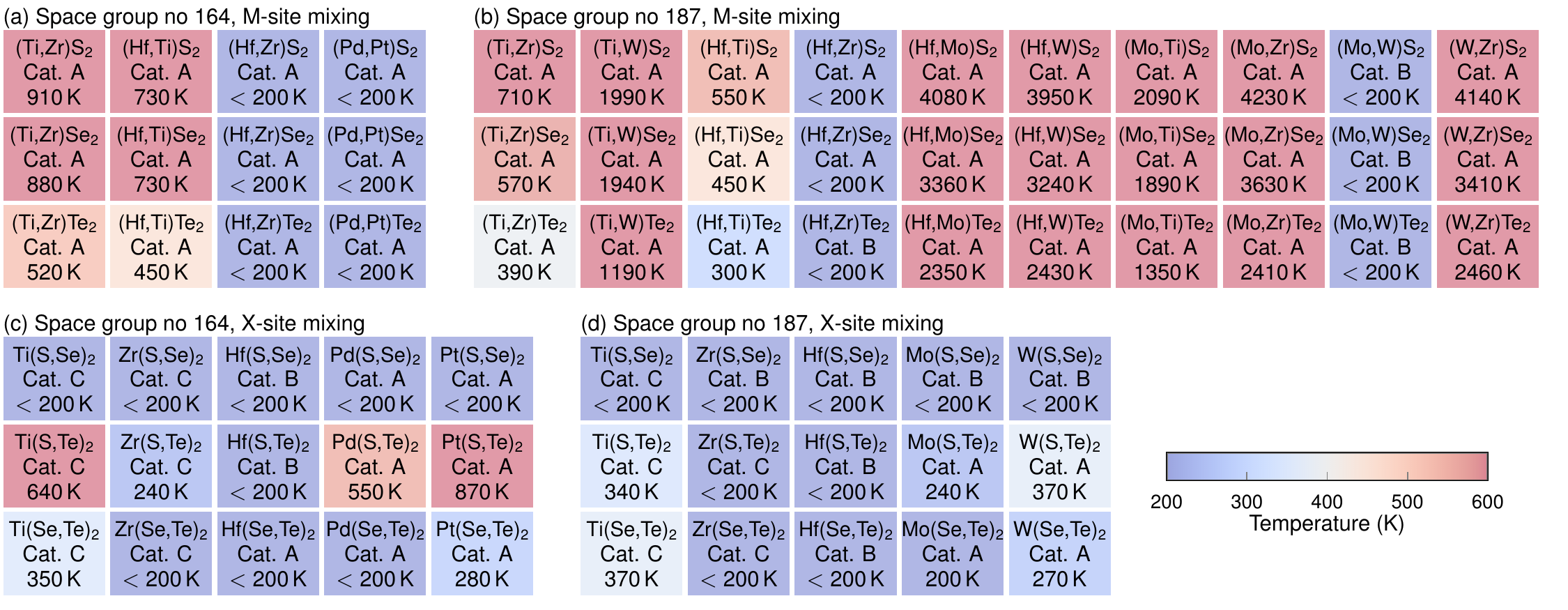}
    \caption{
    Mixing behavior and critical temperatures for all 72 systems studied here.
    Each box indicates predicted mixing behavior close to zero temperature: \emph{category A} for non-mixing systems that phase separate (lateral heterostructures), \emph{category B} for systems that form alloys with long-range order, and \emph{category C} for systems that form Janus monolayers.
    Temperatures indicate the predicted critical temperatures (rounded to nearest \SI{10}{\kelvin}) at which this quality is lost at 50\% concentration.
    Colors are proportional to the critical temperatures.
    }
    \label{fig:critical-temperature}
\end{figure*}

\subsubsection{Category A: Phase separation into boundary phases}
\label{sect:category-a}

\paragraph*{M-site mixing with elements from different groups.}
Among the 38 M-site alloys that are non-mixing at zero temperature, 18 have critical temperatures above approximately \SI{1200}{\kelvin} (\autoref{fig:critical-temperature}a,b), which suggests that these combinations should be difficult to stabilize if at all.
This group comprises all alloys that mix species from different groups of the periodic table on the M-site such as Hf and Mo or W and Zr.
According to the Hume-Rothery rules, well established for metallic alloys \cite{HumMabCha34}, this behavior can be rationalized by the electronegativity differences.
The group 4 elements exhibit electronegativities of 1.3 to 1.5 on the Pauling scale \cite{All61} whereas the group 6 transition metals have values from 2.2 to 2.4 (\autoref{stab:electronegativities}).
This indicate that the bonding in \glspl{tmd} involving group 4 species is much more ionic than in their counterparts from group 6.
For some of these alloying systems, e.g., \ce{Zr_xMo_{1-x}_2} and \ce{Ti_{x}Mo_{1-x}Te2}, there is even a sign difference in the \glspl{bec} (\autoref{stab:boundary-phase-features}), which is a further indication of the incompatibility of the mixing species.
The critical temperature of alloys involving combinations of Zr or Hf with Mo or W is only weakly dependent on the specific transition metal combination.
Rather the critical temperature is primarily determined by the chalcogen species with S and Te-containing systems exhibiting the highest and lowest critical temperatures, respectively.

\paragraph*{M-site mixing with elements from the same group.}
The remaining 20 non-mixing M-site alloys have critical temperatures below approximately \SI{900}{\kelvin} (\autoref{fig:critical-temperature}a,b).
The 9 systems that do not contain Ti even have critical temperatures below \SI{200}{\kelvin} and should thus be readily miscible.
The smaller miscibility in Ti-containing systems can be explained by the much larger difference in lattice parameter between Ti and Zr/Hf-based \glspl{tmd} (\autoref{stab:boundary-phase-features}), which is mirrored by the electronegativity differences (\autoref{stab:electronegativities}).
In fact, as will be discussed in \autoref{sect:models}, the lattice parameters of the boundary phases are in general strong predictors for miscibility.
As in the case of M-site mixing with elements from different groups, the critical temperatures decrease monotonically from S via Se to Te.

\paragraph*{X-site mixing.}
The critical temperatures for non-mixing X-site alloys are generally very low, with the sole exception of the \ce{MS_{2x}Te_{2(1-x)}} alloys with Pd and Pt that still exhibit values of \SI{550}{\kelvin} and \SI{870}{\kelvin}, respectively (\autoref{fig:critical-temperature}c,d).
Generally, the highest critical temperatures are obtained for S--Te alloys, which can be attributed to the large lattice mismatch between the \ce{MS2} and \ce{MTe2} boundary phases.
We note that according to the Hume-Rothery rules a 15\% difference or more in the covalent radii should indicate immiscibility.
Yet, for the present systems the covalent radius is not a good predictor for how well Te mixes with other chalcogen species.
This is confirmed by the analysis in \autoref{sect:models}.

\subsubsection{Category B: Systems with in-plane ordering}
\label{sect:category-b}

The systems in this category include both M-site and X-site mixing that all exhibit an intermediate phase at 50\% with in-plane ordering (\autoref{fig:order-parameters-examples}b).

There are 4 M-site alloys, namely H-\ce{Mo_{x}W_{1-x}X2} with X=S,Se, and Te as well as H-\ce{Hf_{x}Zr_{1-x}Te2} (\autoref{fig:critical-temperature}b), and 8 X-site alloys, including H-\ce{MS_{2x}Se_{2(1-x)}} with M=Mo, W, and Zr as well as five Hf-based systems (\autoref{fig:critical-temperature}c,d).
The only X-site Hf-based alloy that does not exhibit this type of long-range order is T-\ce{HfSe_{2x}Te_{2(1-x)}}.

Generally, the ordering tendency is weak and the critical temperatures are all below \SI{200}{\kelvin}.
For almost all practical purposes these systems can be therefore considered as fully miscible.

\subsubsection{Category C: Systems with out-of-plane (Janus-type) ordering}
\label{sect:category-c}

The systems in this category have the general composition MXX' where one chalcogen species (X) occupies the lattice sites above the transition metal and the other chalcogen species (X') occupies the sites below.

All Ti and Zr-based X-site alloys in both T and H-structure types with the single exception of H-\ce{ZrSSe} have a Janus ground state at 50\% composition.
The critical (order-disorder) temperature was obtained from \gls{mc} simulations by following the difference in composition in each individual layer.
The thus obtained critical temperatures are above room temperature for T-TiSTe, T-TiSeTe, H-TiSTe, and H-TiSeTe with value as high as \SI{640}{\kelvin} for TiSTe, indicating remarkable stability.
To the best of our knowledge, these Janus monolayers have not been synthesized yet.
On the other hand, we find MoSSe, which has in fact been synthesized  \cite{LuZhuXia17}, to be \emph{thermodynamically} unstable in the present study (\autoref{sfig:emix-sg187-B}).
While the Janus monolayer is in fact the isomolar alloy with the highest energy, its energy is only about \SI{13}{\milli\electronvolt/site} equivalent to about \SI{150}{\kelvin} and thus the driving force for decomposing such a structure is very small.
This supports the notion that the growth conditions play a crucial role in enabling the formation of these structures.

It is instructive to consider the factors that lead to thermodynamically stable Janus monolayers by comparing the prototypical cases of H-MoSSe and T-ZrSSe.
The basic parameters of the boundary systems in these two cases are very similar (\autoref{stab:boundary-phase-features}).
The most striking difference is observed the \glspl{bec}, which differ qualitatively.
While in H-MoSSe the elements of the \gls{bec} tensor are negative as has been shown for the boundary phases also in other studies \cite{PikVanDew17}, in T-ZrSSe these elements are positive.
This finding is supported in \autoref{sect:models}, where we construct a classifier for phase behavior, which reveals that the \glspl{bec} are important indicators for the appearance of thermodynamically favorable Janus-structures.

\subsubsection{Comparison with experiment}

Experimentally very little data is available with regard to the thermodynamic properties of \gls{tmd} alloys, which can be attributed to the considerable difficulties associated with such measurements.
It has been observed that phase separation in \ce{MoS_{2x}Te_{2(1-x)}} occurs mainly on the Te-rich side \cite{HibYamHas20}, which was suggested to be due to formation of the distorted 1T' phase. 
In the present study, we have only considered isostructural phase diagrams but nonetheless we observe an asymmetric binodal (\autoref{sfig:emix-sg187-B}).
At 80\% S the alloy should phase-separate at around \SI{275}{\kelvin}, whereas for 20\% that temperature rises to around \SI{325}{\kelvin}.
This suggests that the experimentally observed asymmetry could be already be explained by the thermodynamics of the isostructural system.

For the Mo$_{1-x}$W$_x$S$_2$ alloy, random mixing at room temperature for crystals grown with chemical vapour transport was recently shown \cite{XiaLohVin21}, which is what is expected from the present determination of the critical temperature for the order-disorder transition of \SI{30}{\kelvin}. 

The present study also has potential implications for the understanding of the formation and stability of two-dimensional lateral heterostructures.
It is apparent that the existence of a miscibility gap is not a prerequisite for forming lateral heterostructures with sharp interfaces.
Such a configuration has, for example, been realized between \ce{MoS2} and \ce{WS2} \cite{SusKocKut17, SahMemXin18}, for which both the present and earlier calculations predict a clear preference for mixing (\autoref{sfig:emix-sg187-A-2}).
The small negative mixing energy of down to \SI{-6.5}{\milli\electronvolt/site} in combination with the mixing entropy should imply that such interfaces roughen and vanish at sufficiently temperatures when sufficient kinetic activation enables the system to achieve a thermodynamically more favorable configuration.

On the other hand, for phase separating systems, which can be synthesized in mixed form under non-equilibrium growth conditions, high-temperature annealing, may be used to construct temperature stable lateral heterostructures \cite{SusHacYan18}.  In this context, \autoref{fig:critical-temperature} and \autoref{fig:vbm} provide guidance as to which heterostructures can and cannot be manufactured using annealing.
For example, the proposed lateral heterostructure of T-ZrS$_2$ and T-HfS$_2$ \cite{YuaYuWan18} is likely to require synthesis techniques beyond thermal annealing due to the low critical temperature of \SI{30}{\kelvin} for the T-Zr$_x$Hf$_{1-x}$S$_2$ alloy.

\subsubsection{Comparison with previous modeling studies}

The critical temperatures of the \ce{MS_{2x}Te_{2(1-x)}} and \ce{MSe_{2x}Te_{2(1-x)}} alloys with M=Mo or W has been computed before in Ref.~\citenum{KanTonLi13}, which used a mean field approximation of the configurational entropy and the PBE exchange-correlation functional.
Here, we go beyond a mean field approximation by using \gls{mc} simulations that explicitly sample the compositional configuration space and use a more advanced exchange-correlation functional that can be expected to provide a better description of these systems \cite{LinErh16}.
Overall we find qualitative agreement, and in the case of the Se--Te alloys there is a satisfying quantitative agreement as well.
We predict that the critical temperature is \SI{270}{\kelvin} (\SI{200}{\kelvin}) to be compared with \SI{360}{\kelvin} (\SI{280}{\kelvin}) for the W (Mo) based alloy.
For the S--Te alloys, however, we find considerable lower critical temperatures of \SI{370}{\kelvin} (\SI{240}{\kelvin}) to be compared with \SI{690}{\kelvin} (\SI{490}{\kelvin}) for the W (Mo) based alloy. 
It is also apparent that the present calculations yield a more asymmetric phase diagram, as already noted above.
This suggests that the explicit treatment of the configurational entropy and the exchange-correlation are likely to play a role here.

It has been previously estimated based on a mean field approximation of the configurational entropy that the critical temperature for the order-disorder transition in Mo$_{1-x}$W$_x$S$_2$ alloy for x = 1/3 and x = 2/3 is around \SI{40}{K} \cite{TanWeiLiu2017}, which is very close to the value of \SI{30}{K} obtained here.

\subsection{Models for predicting alloying behavior}
\label{sect:models}

\begin{figure}
\centering
\includegraphics[scale=0.9]{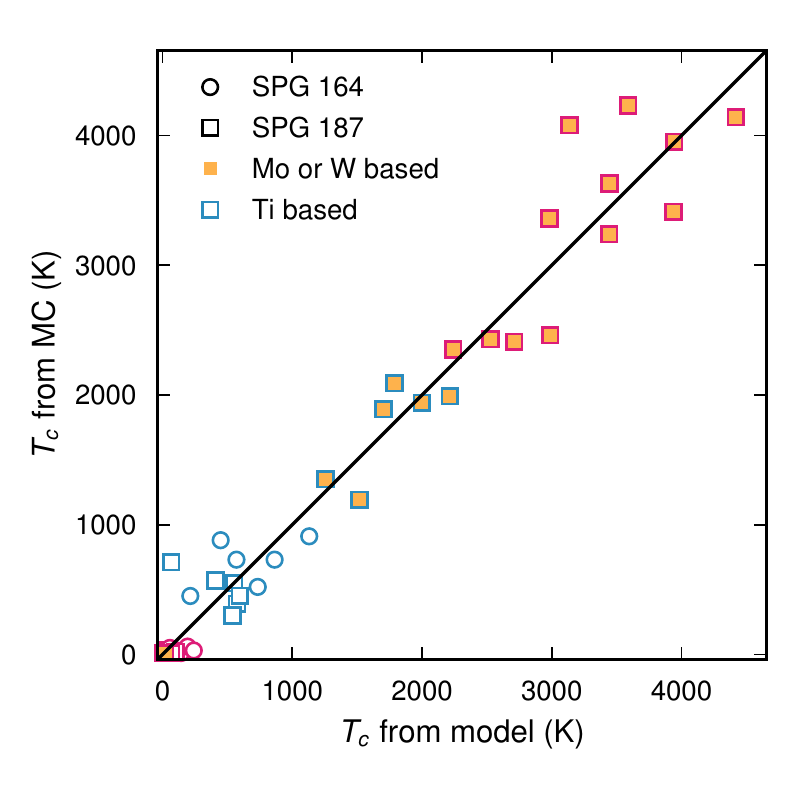}
\caption{
    Parity plot of the critical temperature for M-site alloys as computed via \gls{mc} simulations and the critical temperature model based on the descriptor in Eq.~\eqref{eq:model}.
}
\label{fig:model}
\end{figure}

In the previous sections we already indicated how certain properties of the boundary phases can serve as qualitative predictors for the properties of the mixed system.
Now we aim to analyze such relations more quantitatively.
In the case of \emph{metallic} (binary) alloy the Hume-Rothery rules are well established as a set of basic conditions under which one can expect mixing.
According to these (qualitative) rules mixing ought to be possible if (\emph{i}) the relative difference of the covalent radii of the components is less than 15\% and the components have (\emph{ii}) similar electronegativity as well as (\emph{iii}) the same electronic valency.
Even though these rules have been formulated for metallic systems, they can still partially explain the mixing behavior in several of the \gls{tmd} alloys considered here.
The Hume-Rothery conditions are, however, based on the properties of the free atoms and as such do not account for any many-body interactions in the boundary phases.
For example, in the case of \ce{MSTe} alloys, the covalent radii of the chalcogens exhibit a 30\% difference yet the critical temperature ranges from around \SI{200}{\kelvin} to 
around \SI{1000}{\kelvin} indicating that the contributions from interactions in the boundary phases are crucial for the alloying behavior.

It can be a daunting task to identify the features of the boundary phases that are the most suitable for predicting alloying ability.
Thankfully there are machine learning methods that are very well suited for this task.
For this purpose, here, we employ the \gls{sisso} approach that provides an automatized procedure for feature selection to construct predictors for the critical temperature as well as the alloy category (\autoref{fig:critical-temperature}) \cite{OuyCurAhm18}.
To this end, we considered the following properties of the boundary phases:
lattice parameters ($a_0$), covalent radii, \glspl{bec}, electronegativities, unit cell areas ($A$), sheet modulus ($B$), ionization potential (IP), electron affinity ($\chi$), and work function.
Original features were constructed as the magnitude of the difference between the boundary phases.
Since we found the the sign of the \glspl{bec} to be relevant in the case of the Janus monolayers (\autoref{sect:category-c}) for the categorization model we constructed a signed \gls{bec} features as
\begin{align}
    \gamma_{\mathrm{BEC}} = \frac{\mathrm{Tr}(\mathbf{\mathcal{Z}}^{\ast}_1)-\mathrm{Tr}(\mathbf{\mathcal{Z}}^{\ast}_0)}{\mathrm{Tr}(\mathbf{\mathcal{Z}^{\ast}_0})},
\end{align}
where $\mathrm{Tr}(\mathbf{\mathcal{Z}}^{\ast}_1) > \mathrm{Tr}(\mathbf{\mathcal{Z}}^{\ast}_0)$ for the two boundary phases indicated with subscripts 0 and 1, respectively.
This quantity provides the information regarding the \glspl{bec}, removes the ambiguity due to large anisotropy of the monolayers, and at the same time contains information regarding the sign of the \glspl{bec}.

\begin{figure}
\centering
\includegraphics{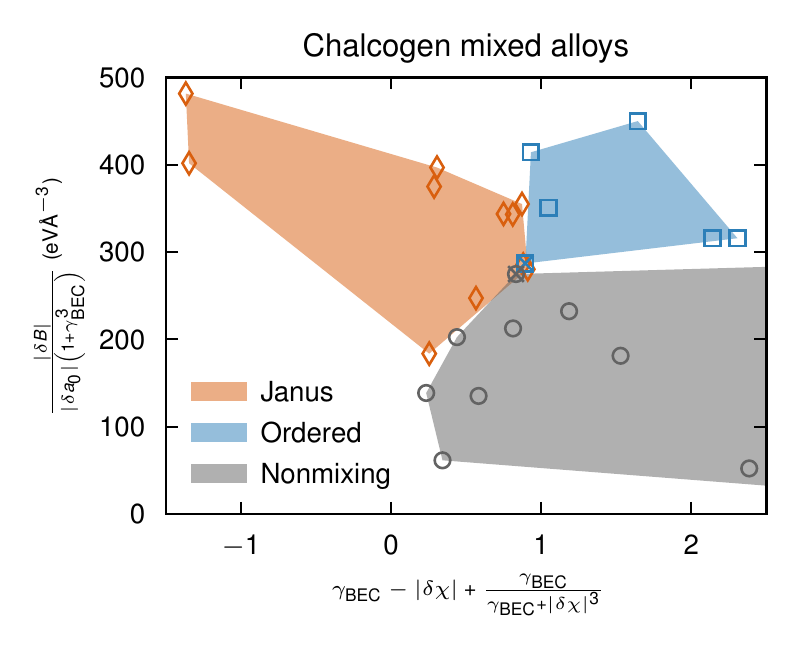}
\caption{
    Classification of X-site alloys into ordered, Janus or non-mixing ground state using the descriptors in Eqs.~\eqref{eq:classification-descriptor-1} and \eqref{eq:classification-descriptor-2}.
}
\label{fig:classification}
\end{figure}

We note that some of the considered features are highly correlated, e.g., the lattice parameter depends on the character of the bonding in the solid, which is also connected to the \glspl{bec}, electronegativity, and bulk modulus. 

With the exception of only three systems the critical temperatures of the X-site alloys considered here are below \SI{370}{\kelvin}.
We therefore only consider M-site alloys here, for which we seek a linear relationship between between the descriptor to be determined and the critical temperature.
The descriptor that we find via \gls{sisso} is 
\begin{equation} \label{eq:model}
    f = \sqrt{|\delta \mathrm{IP}|} \sqrt[3]{|\delta B| |\delta A|^2},
\end{equation}
which yields a \gls{rmse} of \SI{292}{\kelvin} and a coefficient of determination $R^2$ of 0.956 (\autoref{fig:model}).
The descriptor is not unique and depends on the features considered in the model as well as the set of alloys considered as discussed in the original reference \cite{OuyCurAhm18}.
More crucially it, however, provides clear indications as to which boundary phase features to consider when rationalizing alloying behavior.
Specifically, the descriptor here emphasizes that ionization potential (or the valence band edge position) as well sheet modulus and unit cell area are important parameters.

In addition we constructed descriptors to enable a categorization of X-site alloys according to the classification introduced in \autoref{sect:phase-diagrams}.
The resulting model has a 93\% success rate in partitioning the different categories of mixing behaviour (\autoref{fig:classification}) and uses two descriptors.
The first descriptor (abscissa in \autoref{fig:classification}) obtained via \gls{sisso} is given by
\begin{align}
    d_1 = \gamma_{\mathrm{BEC}} - |\delta \chi| + \frac{\gamma_{\mathrm{BEC}}}{\gamma_{\mathrm{BEC}} + |\delta \chi|^3}
    \label{eq:classification-descriptor-1}
\end{align}
It is dimensionless and contains features directly related to bonding ($\gamma_{\mathrm{BEC}}$ and $|\delta \chi|$).
The second descriptor (ordinate axis in \autoref{fig:classification}) is
\begin{align}
    d_2 = \frac{|\delta B|}{|\delta a_0|(1 + \gamma_{\mathrm{BEC}}^3)}
    \label{eq:classification-descriptor-2}
\end{align}
and has units of pressure.
Unfortunately, neither of these two descriptors are amenable to a clear physical interpretation.

\begin{figure*}
    \centering
    \includegraphics[width=\linewidth]{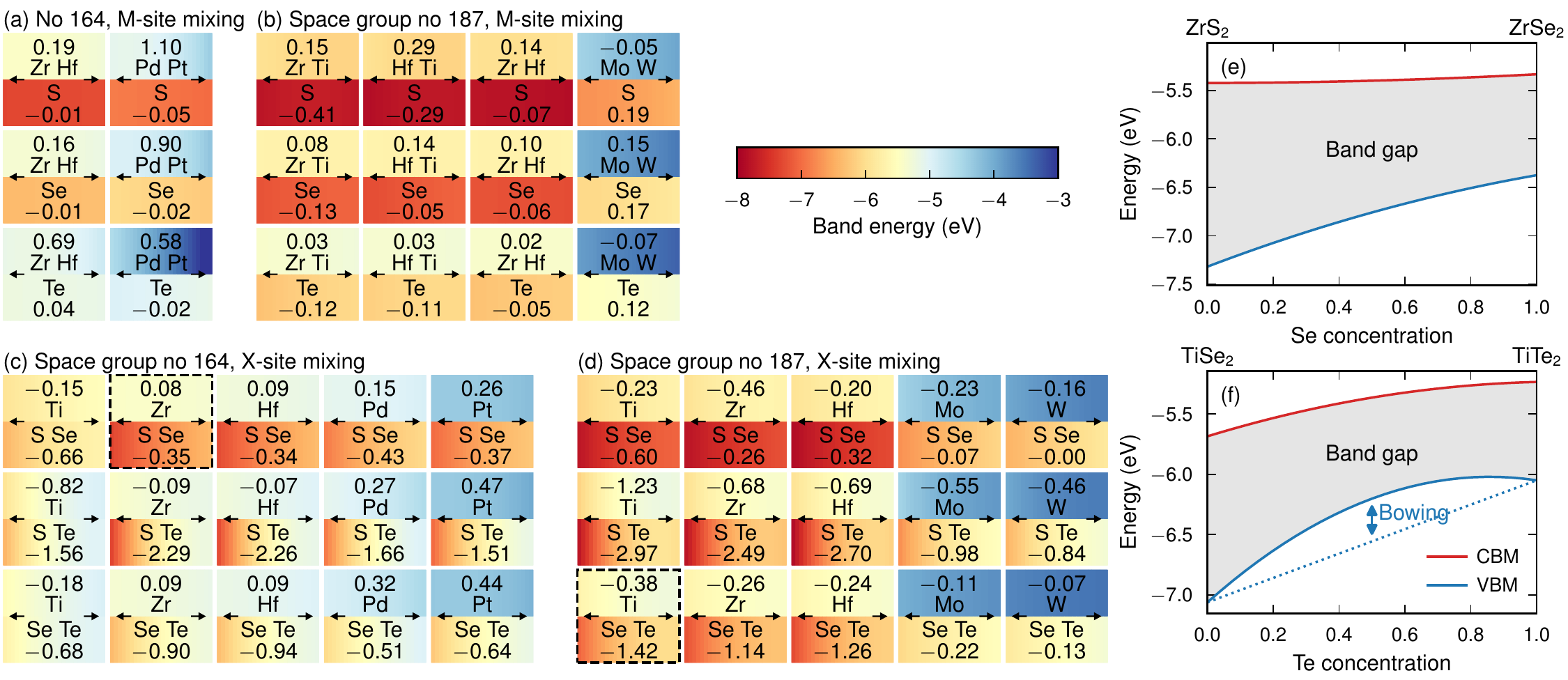}
    \caption{
        (a--d) Conduction band and valence band edge positions with respect to vacuum for H and T-type M and X-site alloys.
        Each box contains a heat map of the position of the conduction band (top) and valence band (bottom).
        The bowing parameters (see Eq. \eqref{eq:bowing-parameter}) in units of eV for the conduction band (valence band) are indicated in the top (bottom) of the box.
        (e--f)~Conduction band minimum (CBM) and valence band maximum (VBM) as a function of concentration in (e) T-ZrS$_{2(1-x)}$Se$_{2x}$ and (f) H-TiSe$_{2(1-x)}$Te$_{2x}$.
        The bowing parameter is a measure of the deviation of the true band position from a linear interpolation of the band positions of the boundary phases as indicated by the dotted, blue line in (f).
    }
    \label{fig:vbm}
\end{figure*}

Two systems are incorrectly categorized.
\ce{HfS_{2x}Se_{2(1-x)}} is categorised as both a Janus and an ordered system, while in reality it is an ordered system yet the Janus structure is only $\leq\!\SI{5}{\milli\electronvolt\per\formulaunit}$ than other configurations.
\ce{PdS_{2x}Se_{2(1-x)}} (categorised as both non-mixing and Janus) on the other hand is a non-mixing system with a critical temperature of \SI{80}{\kelvin}, but the Janus structure is \SI{30}{\milli\electronvolt\per\formulaunit} higher in energy than other configurations.

\subsection{Electronic structure}
\label{sect:electronic-structure}

Following the analysis of the thermodynamic properties, we carried out a systematic evaluation of the valence and conduction positions as a function of composition.
From this analysis we excluded systems with very high critical temperature, leaving us with 48 systems (\autoref{fig:vbm}).

To first order the band edge positions change linearly with composition between the boundary phases, a behavior that is often referred to as Vegard's law.
The deviation from this dependence is commonly described by the bowing parameter $b$, which is defined via
\begin{align}
    \varepsilon(x) = (1-x) \varepsilon(x=0) + x \varepsilon(x=1) - b x (1-x),
    \label{eq:bowing-parameter}
\end{align}
where $\varepsilon$ represents the concentration dependent quantity in question.
Note that according to this (standard) definition $-b/4$ corresponds to the deviation from a linear interpolation at $x=0.5$.

Band edge bowing parameters for MX$_{2x}$X'$_{2(1-x)}$ for M=Mo,W, and X,X'=S,Se,Te were already reported in Ref.~\citenum{KanTonLi13}.
The current results are overall in good agreement with these data, albeit slightly smaller in magnitude (\autoref{stab:bowing}).

The general trend for X-site alloys is that the valence band bowing is negative and increasing in magnitude in the order S$_{2x}$Se$_{2(1-x)}$,  Se$_{2x}$Te$_{2(1-x)}$, S$_{2x}$Te$_{2(1-x)}$ (\autoref{fig:vbm}c,d).
Especially for the S$_{2x}$Te$_{2(1-x)}$ alloys, the magnitude of the bowing parameter can exceed values of \SI{1}{\electronvolt}.
For the conduction band the variations are generally smaller and there is no apparent trend that extends through all \glspl{tmd}.
For example, for T-type Zr and Hf-based \glspl{tmd} the bowing parameters for S$_{2x}$Se$_{2(1-x)}$, Se$_{2x}$Te$_{2(1-x)}$, and S$_{2x}$Te$_{2(1-x)}$ are all in range between 0.07 and \SI{0.09}{\electronvolt}.
Furthermore, we find that for X-site alloys, the bowing parameter of the valence band is always negative, and that for H-type X-site alloys, the conduction band bowing parameter is always negative.
For T-type X-site alloys, the conduction band bowing parameter can have either sign.
For example, in T-HfS$_{2x}$Se$_{2(1-x)}$, the valence band exhibits a bowing parameter of \SI{-0.34}{\electronvolt} while the conduction band bowing parameter is \SI{0.09}{\electronvolt}.

For T-type M-site alloys, the magnitude of the bowing of the valence band edge is generally small, with the maximal (absolute) value of \SI{-0.05}{\electronvolt} occurring in T-Pd$_x$Pt$_{1-x}$S$_2$ (\autoref{fig:vbm}a,b). The conduction band edge exhibits significantly larger bowing in this class of alloys with values up to \SI{1.10}{\electronvolt} (in T-Pd$_x$Pt$_{1-x}$S$_2$). Finally, for H-type M-site alloys, the valence band bowing is larger in magnitude than for T-type M-site alloys with a maximal magnitude of \SI{-0.41}{\electronvolt} found in H-Ti$_x$Zr$_{1-x}$S$_2$. The largest bowing of the conduction band for the H-type M-site alloys is found in Ti$_{x}$Hf$_{1-x}$S$_2$ with a value of \SI{0.29}{\electronvolt}. 

The valence band and conduction band variations that are possible in the considered alloys are illustrated in \autoref{fig:vbm}.
Specifically, we find that the variation of the valence band can be very large in \ce{MS_{2x}Te_{2(1-x)}}, e.g., in HfS$_{2x}$Te$_{2(1-x)}$ the difference between the boundary phases is \SI{1.7}{\electronvolt}.
In general the valence band position increases in the series S--Se--Te.
The typical behavior is illustrated for T-\ce{ZrS_{2x}Se_{2(1-x)}} (\autoref{fig:vbm}e) and H-\ce{TiSe_{2x}Te_{2(1-x)}} (\autoref{fig:vbm}f).

It has been noted before that the magnitude of the valence band bowing parameter increases in the order S$_{2x}$Se$_{2(1-x)}$, Se$_{2x}$Te$_{2(1-x)}$, and S$_{2x}$Te$_{2(1-x)}$ for H-type Mo and W-based X-site alloys \cite{KanTonLi13}.
Our results show that this trend also holds true for Ti, Hf, and Zr-based alloys that exhibit H symmetry and Zr, Hf, Ti, Pd, and Pt alloys that exhibit T symmetry.
The band edges exhibit in most cases moderate bowing with the exception of the X-site alloys \ce{MS_{2x}Te_{2(1-x)}}.
These are the X-site mixed systems that exhibit the largest lattice constant mismatch and largest differences in boundary phase features. Therefore, these alloys exhibit the largest variability of the considered alloys.
The large bowing parameter in these compounds may however complicate band edge tuning since a small variation in composition is associated with a relatively large variation in band edge position.

\section{Conclusions}

In this study, we have provided a comprehensive study of the phase diagrams and electronic structure of monolayer \gls{tmd} alloys.
It has been shown that mixing systems with in-plane order are absent at room temperature but that specifically the Ti-based X-site alloys may exhibit a Janus ground state that remains the most stable structure far above room temperature.
Furthermore, we have shown that M-site alloys with transition metals from different groups in the periodic table are associated with very high critical temperatures ($>\SI{1000}{\kelvin}$) whence it is unlikely or at least very difficult for them to be manufactured.
The band edges exhibit in most cases little bowing with the exception of X-site alloys that include both S and Te, implying that in many systems the band edge positions are relatively well approximated by a linear interpolation of the values of the boundary phases.

\section*{Acknowledgments}
Funding from the Knut and Alice Wallenberg Foundation (2014.0226) as well as the Swedish Research Council (2018-06482, 2020-04935) are gratefully acknowledged.
The computations were enabled by resources provided by the Swedish National Infrastructure for Computing (SNIC) at NSC, C3SE, and HPC2N partially funded by the Swedish Research Council through grant agreement no. 2018-05973.

\end{document}